\documentclass[fleqn,usenatbib]{mnras}
\usepackage{newtxtext,newtxmath}
\usepackage[T1]{fontenc}
\usepackage{ae,aecompl}

\usepackage{graphicx}	
\usepackage{amsmath}	
\usepackage{amssymb}	

\title[X-ray Cavities in NGC 4477]{X-ray cavities in the hot corona of the lenticular galaxy NGC~4477}

\author[Yijia Li, Yuanyuan Su and Christine Jones]{
Yijia Li$^{1,2}$,
Yuanyuan Su$^{2}$, 
Christine Jones$^{2}$
\\
$^{1}$School of Astronomy and Space Science, Nanjing University, Nanjing 210046, China\\
$^{2}$Harvard-Smithsonian Center for Astrophysics, 60 Garden Street, Cambridge, MA 02138
}

\date{}

\pubyear{2018}

\begin{document}
\label{firstpage}
\pagerange{\pageref{firstpage}--\pageref{lastpage}}
\maketitle

\begin{abstract}
NGC 4477 is a low-mass lenticular galaxy in the Virgo Cluster, residing at 100\,kpc to the north of M87. Using a total of 116\,ks {\sl Chandra} observations, we study the interplay between its hot ($\sim$0.3\,keV) gas halo and the central supermassive black hole. A possible cool core is indicated by the short cooling time of the gas at the galaxy centre. We identify a pair of symmetric cavities lying 1.1\,kpc southeast and 0.9\,kpc northwest of the galaxy centre with diameters of 1.3\,kpc and 0.9\,kpc, respectively. We estimate that these cavities are newly formed with an age of $\sim$4\,Myr. No radio emission is detected at the positions of the cavities with the existing VLA data. The total energy required to produce the two cavities is $\sim$$10^{54}$\,erg, at least two orders of magnitude smaller than that of typical X-ray cavities. NGC 4477 is arguably far the smallest system and the only lenticular galaxy in which AGN X-ray cavities have been found. It falls on the scaling relation between the cavity power and the AGN radio luminosity, calibrated for groups and clusters. Our findings suggest that AGN feedback is universal among all cool core systems. Finally, we note the presence of molecular gas in NGC~4477 in the shape of a regular disk with ordered rotation, which may not be related to the feedback loop. 
\end{abstract}


\section{Introduction}
\label{sec:1}

Over the last few decades, strong observational evidence has emerged supporting the co-evolution of supermassive black holes and their host galaxies (see \citealt{2013ARA&A..51..511K} and \citealt{2014ARA&A..52..589H} for reviews). Best known is the tight correlation between the black hole mass and the velocity dispersion of the host galaxy's bulge (see \citealt{2013ARA&A..51..511K}). Feedback from the active galactic nuclei (AGN) may be important in this co-evolution \citep[e.g.][]{2012ARA&A..50..455F}. Accretion onto the AGN produces radiation, winds and jets, which interact with the interstellar medium (ISM).
AGN feedback provides a promising solution to the cooling problem \citep[e.g.][]{2007ARA&A..45..117M}. In the classical radiative cooling flow model, the hot gas in the central region of galaxy groups and clusters cools by radiating in X-rays. Being compressed by the surrounding gas, cooled gas flows inward towards the central black hole, causing prodigious star formation in cool core systems \citep{1994ARA&A..32..277F}. But the generally observed low star formation rates imply a gas cooling rate much smaller than expected. It was therefore proposed that cooling gas is reheated by AGN feedback. It remains debated how the gas state influences the activity of AGN and how energy from the AGN is transported. It is also unclear whether the mechanism that reheats the cooling gas in massive elliptical galaxies, e.g, the brightest cluster galaxies (BCG), can be applied to small galaxies. 

AGN outbursts inject jet power into the ambient hot gas corona and produce X-ray cavities, characterized by decrements in the X-ray surface brightness. These cavities are usually filled with relativistic particles (see \citealt{2010ApJ...712..883D} and \citealt{2016ApJS..227...31S} for systematic searches for cavities). By inflating relativistic bubbles, the outburst energy can be deposited into the hot gas.
By measuring the cavity size and its distance to the nucleus, we can calculate the amount of energy released by the supermassive black hole and estimate its duty cycle.

X-ray cavities have been observed in bright elliptical galaxies at the centres of clusters and groups as well as in some individual elliptical galaxies, e.g. 
MS0735.6+7421 \citep{2005Natur.433...45M}, NGC~1399 \citep{2017ApJ...847...94S}, and M84 \citep{2008ApJ...686..911F}.
Systematic studies of X-ray cavities suggest that $\sim$70 per cent of cool-core clusters harbor cavities \citep{2006MNRAS.373..959D}, and this rate drops to $\sim$50 per cent for groups \citep{2010ApJ...712..883D} and $\sim$25 per cent for individual elliptical galaxies \citep{2009AIPC.1201..198N}. 
Lenticular galaxies are regarded as a transitional phase between late-type galaxies and elliptical galaxies and may carry important information of galaxy evolution. To the best of our knowledge, no cavities have been found in lenticular galaxies.

NGC 4477 is a nearby SB0(s) galaxy in the Virgo Cluster, residing at 100\,kpc to the north of M87. The {\sl XMM-Newton} image of NGC~4477 is shown in Figure~\ref{fig:1}. The hot gas corona of NGC~4477 is more extended northwest of the galaxy, likely being stripped as NGC~4477 travels through the intracluster medium (ICM) of Virgo. The current motion of NGC~4477 is likely to be in the plane of the sky given that its radial velocity differs from that of M87 by only $\sim$50\,km\,s$^{-1}$ (NASA/IPAC Extragalactic Data base (NED)).
We assume the distance to NGC~4477 is the same as that of Virgo, $D_L=16.7$\,Mpc (1\,arcsec = 0.08\,kpc; \citealt{2007ApJ...655..144M}).
The central black hole in NGC 4477 has a mass of $10^{7.55 \pm 0.3}$\,M$_{\sun}$ \citep{2012MNRAS.419.2497B}. 
It is categorized as a Seyfert 2 galaxy
with a star formation rate surface density of $10^{-1.13 \pm 0.33}$\,M$_{\sun}$\,yr$^{-1}$\,kpc$^{-2}$\,\citep{2014MNRAS.444.3427D}. 
Its radio emission is not detected by the VLA FIRST at 20 cm, with an upper flux limit of 0.72\,mJy \citep{2017MNRAS.464.1029N}, but the NGC~4477 nucleus is marginally detected at 6 cm\citep{2001ApJS..133...77H}.

In this work, we study the hot ISM of NGC~4477 and its interaction with the central AGN using deep {\sl Chandra} observations. Section 2 is devoted to the observations and data reduction. Results on the global gas properties and the cavities are presented in Section 3. We discuss possible shock fronts in Section 4.1, the cavity heating mechanism in Section 4.2, properties of the nucleus in Section 4.3, and the cold gas content in Section 4.4. We summarize our findings in Section 5.

\section{Observations and Data Reduction}
\label{sec:2}

\subsection{{\textit Chandra} Observation of NGC~4477}
\label{sec:2.1}
We include all three {\sl Chandra} observations of NGC~4477 as listed in Table~\ref{tab:1}.
Only one observation (ObsID 9527), with an exposure time of 38\,ks, was targeted on NGC~4477; its aim point is on the back-side illuminated ACIS-S3 CCD. 
The other two observations are focused on a background cluster XMMU J1230.3+1339 at z = 0.975 (ObsID 11736 and 12209) with NGC~4477 on the ACIS-S2 CCD, off axis by 3.67\arcmin. In this paper, we use the combined $\sim$116\,ks observations to study the global gas properties of NGC 4477. When investigating the cavities, edges, and other detailed structures, we primarily utilize the 38\,ks ACIS-S3 observation. A background subtracted and exposure corrected {\sl Chandra} image (ObsID 9527) in the 0.3--2.0\,keV energy band is shown in Figure~\ref{fig:2}-left. It reveals two cavities along an east-west axis, surface brightness rims outside the cavities, and two bright point sources at the centre. By comparing the surface brightness of the two cavities and that of an annular region at the same radius, we estimate the significance of the detection of cavities to be $\sim$11$\sigma$ for the southeast cavity and $\sim$7$\sigma$ for the northwest cavity.

\begin{figure}
  \begin{tabular}{c}
    \includegraphics[width=\columnwidth]{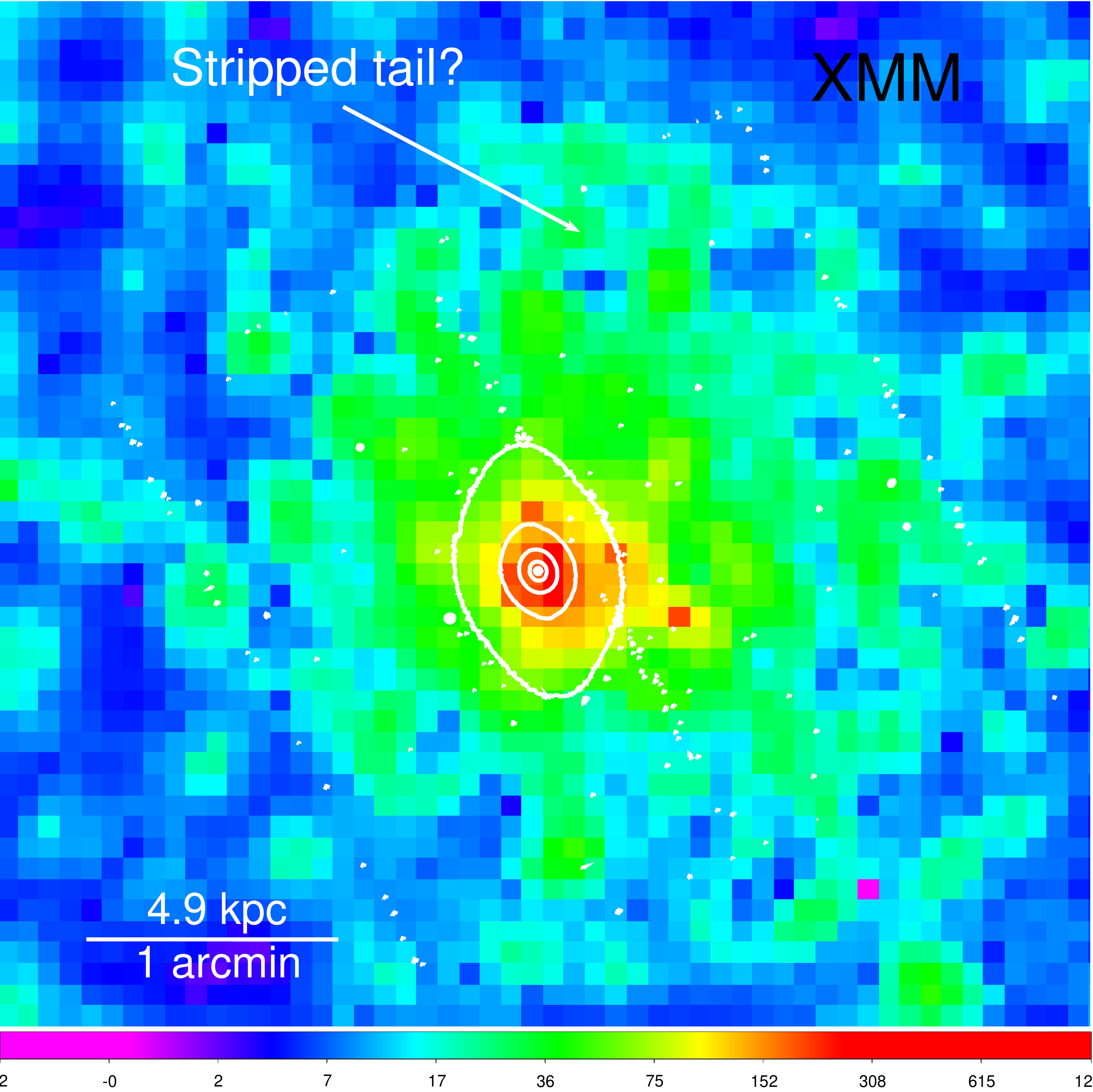}
  \end{tabular}
    \caption{{\sl XMM-Newton} image in the 0.3--2.0\,keV energy band, in units of cts\,s$^{-1}$\,deg$^{-2}$, with the {\sl Hubble Space Telescope} ({\sl HST}) contours (white) overlaid. 10 levels of contours were created for the flux limits 0.15 electron\,s$^{-1}$ to 500 electron\,s$^{-1}$ at a smoothness of 10. The image demonstrates hot gas emission extended to a radius of $\sim$7\,kpc.} 
    \label{fig:1}
\end{figure}

\begin{table*} 
	\centering
	\caption{{\sl Chandra} observations of NGC 4477.}
	\label{tab:1}
	\begin{tabular}[width=\columnwidth]{ccccc}
		\hline
		ObsID & Exposure$^a$ (ks) & Instrument Used & On-axis Target & Start Date\\
		\hline
		9527 & 37.68 & ACIS-S3 & NGC 4477 & 2008-04-27 \\ 
        11736 & 58.06 & ACIS-S2 & XMMU J1230.3+1339 & 2010-04-30 \\ 
        12209 & 19.91 & ACIS-S2 & XMMU J1230.3+1339 & 2010-05-02 \\
		\hline
\multicolumn{5}{p{2.75in}}{$^a$ Exposure time after background flare filtering.}\\
\end{tabular}
\end{table*}

\begin{figure*}
    \includegraphics[width=\columnwidth]{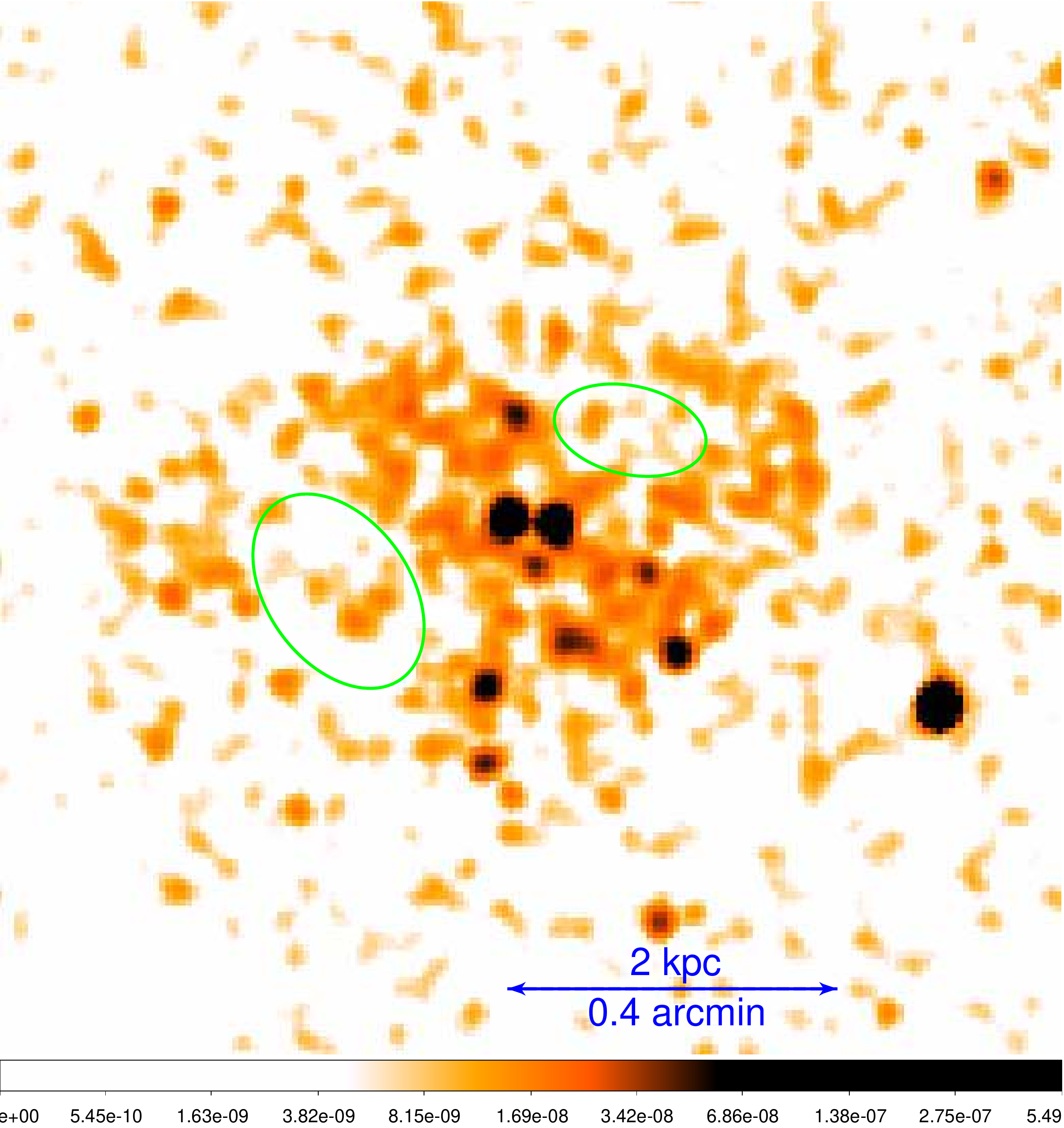} 
    \includegraphics[width=\columnwidth]{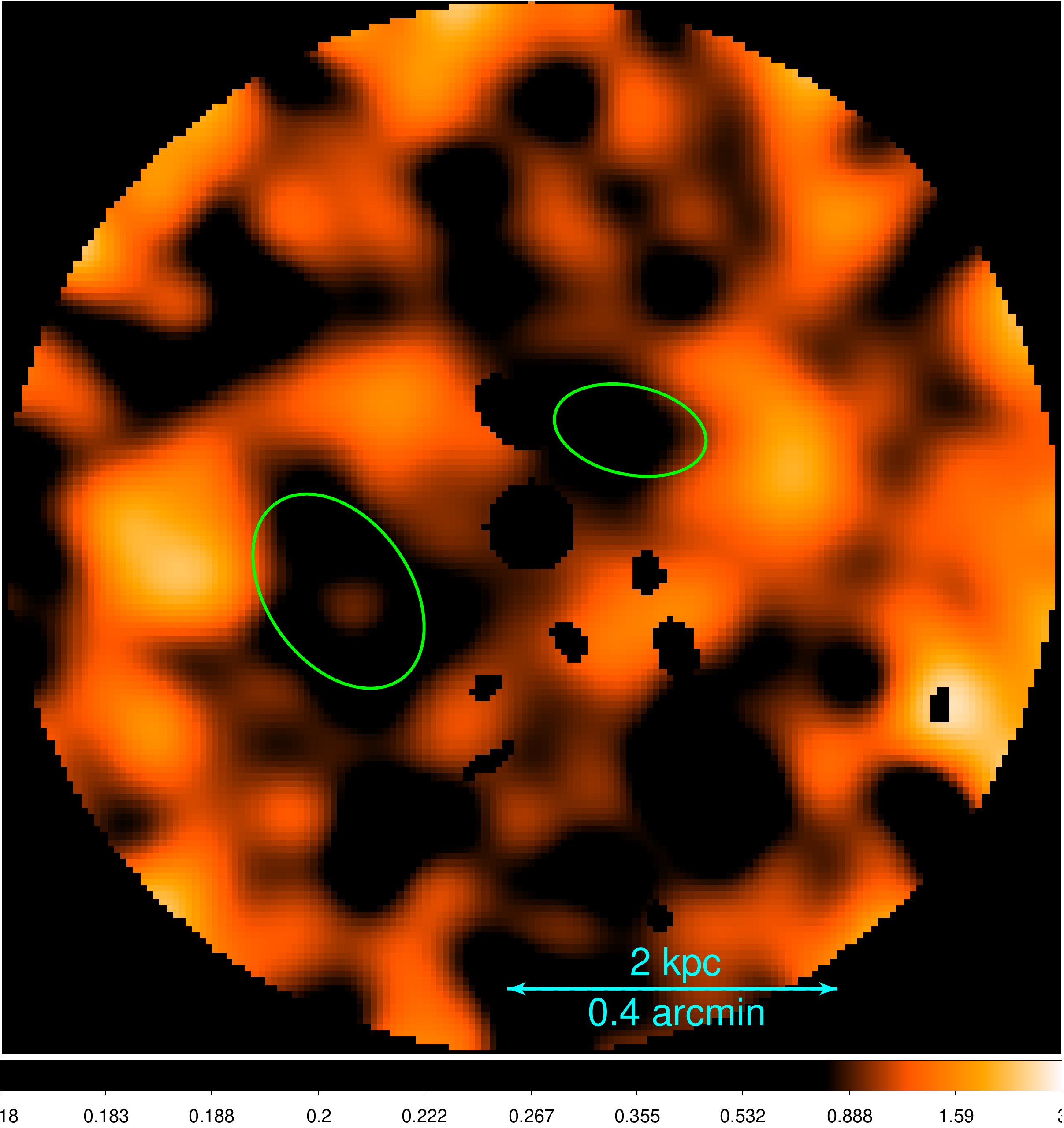}\\
    \caption{ Left: 0.3--2.0\,keV smoothed image of ObsID 9527, in units of photon\,cm$^{−2}$\,s$^{-1}$, with a blank-sky background subtracted. Right: residual image, created by the left {\sl Chandra} image divided a double-$\beta$ profile model, with point sources masked and the centre not fitted. The images are matched. Two X-ray cavities are indicated with green ellipticals.}
    \label{fig:2}
\end{figure*}

\subsection{Data Reduction and Spectrum Analysis}
\label{sec:2.2}
We reduce the {\sl Chandra} data using {\sc CIAO} 4.9 and {\sc CALDB} 4.7.4. Level 1 event files were reprocessed using CHANDRA\_REPRO. Background flares were removed. Identification of point sources were performed for each observation using WAVDETECT with images in the 0.5--7.0\,keV energy band. 
Readout events were removed from both imaging and spectral analyses. All spectra were extracted using SPECEXTRACT. The instrumental and astrophysical background was approximated by the blank-sky fields available in the {\sc CALDB}. Levels of blank-sky background were scaled by the count rates in the 9.5--12.0\,keV energy band relative to the observations. Point sources were omitted in the spectral analysis. All spectra were grouped to a minimum of one count per channel and C-statistics in {\sc XSPEC} 12.9.1 were used for the spectral analysis. The spectral fit of the hot gas was restricted to the 0.5-5.0\,keV energy band. 
 
We extracted spectra from four consecutive annuli to probe the hot gas properties of NGC~4477. Each annulus region contains at least 500 net counts.
We adopted a galactic absorption of $N_H$ = $2.44 \times {10}^{20}$\,cm$^{-2}$ \citep{2005A&A...440..775K}. 
The absorbed hot gas component was modeled with {\tt phabs} $\times$ {\tt apec}. The solar abundance standard of \citet{2006NuPhA.777....1A} was adopted and we fixed the hot gas metallicity to 0.5 solar based on the {\sl XMM-Newton} measurement for NGC~4477 in \citet{2013ApJ...766...61S}. The emission of unresolved low mass X-ray binaries (LMXBs) was modeled with a {\tt power-law} model with an index of 1.6 \citep{2003ApJ...587..356I}. The emission of active binaries (ABs) and cataclysmic variables (CVs) was modeled with a {\tt power-law} model with an index of 1.9 and a {\tt mekal} model with a temperature of 0.5\,keV (\citealt{2008A&A...490...37R}; also see \citealt{2015ApJ...806..156S}). 
We include an additional {\tt apec} component to account for the Virgo ICM emission. 
The best-fit temperature of the ICM is 2.1\,keV.
We fixed the normalizations of the LMXBs component based on the scaling relationship between the X-ray luminosity of LMXBs and the K-band luminosity of early-type galaxies (see \citealt{2006ApJ...653..207D}). The fluxes of the ABs and CVs components were fixed to 10 per cent that of LMXBs (see \citealt{2011ApJ...729...12B}). A large scatter was found in the $L_{\rm LMXB}$--$L_K$ relation. We thus varied the normalizations of the LMXBs component by 20 per cent; its impact on our results is negligible. Deprojection analysis was performed using the mixing model, {\tt\string projct}, in {\sl XSPEC}. Deprojected temperature and density profiles are shown in Figure~\ref{fig:4} in \S3.1.

\section{Results}
\label{sec:3}

\subsection{Global Properties}
\label{sec:3.1}
\begin{figure}
    \includegraphics[width=\columnwidth]{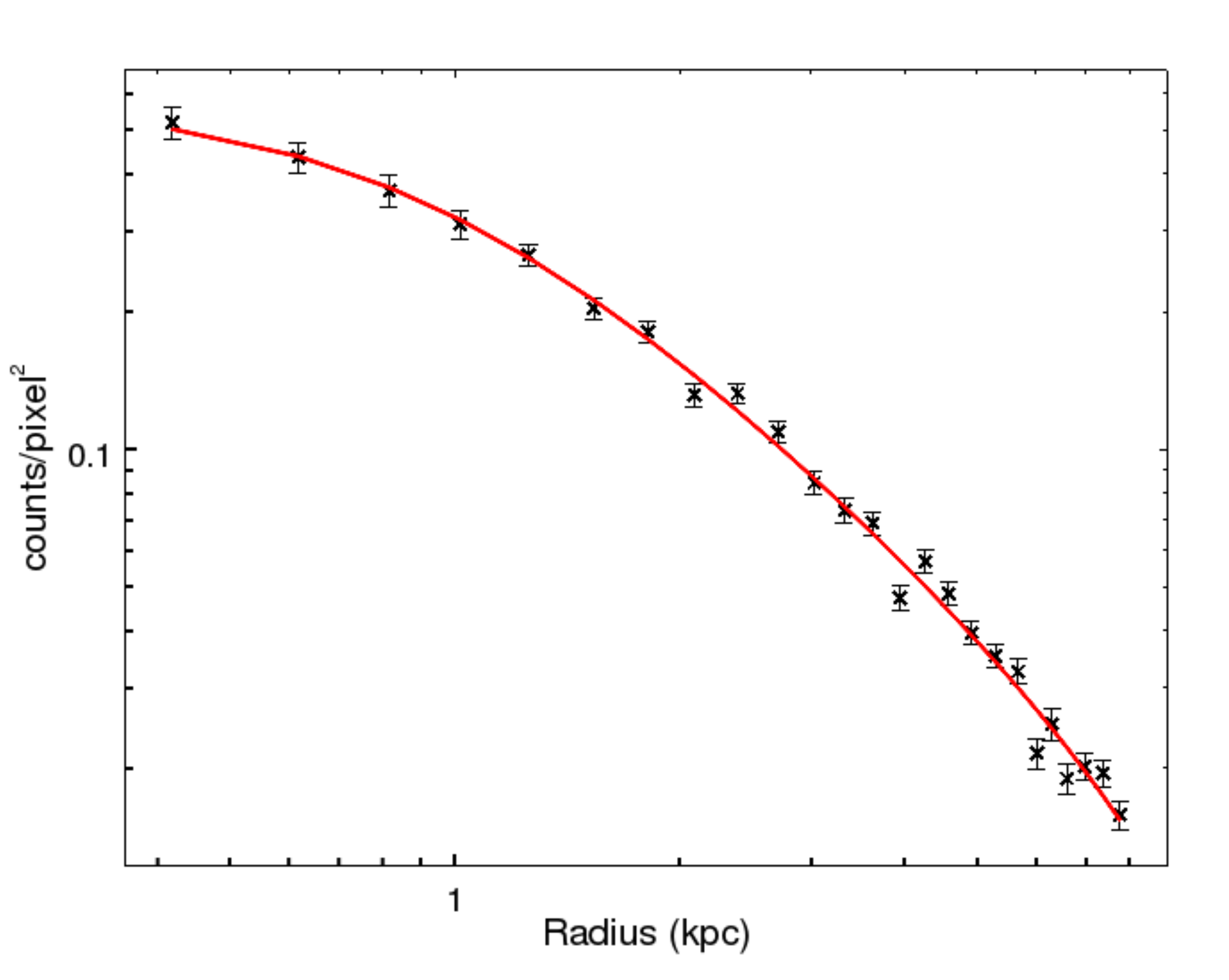} \\
    \caption{Surface brightness profile extracted from the 0.3-2.5\,keV image of all three {\sl Chandra} observations. Point sources have been excluded and the cavity regions are included. The red curve represents the best-fitting $\beta$-model.}
    \label{fig:3}
\end{figure}

As shown in the {\sl XMM-Newton} image in Figure~\ref{fig:1}, the diffuse emission of NGC~4477 extends to a radius of $\sim$7\,kpc.
We obtained an azimuthally-averaged surface brightness profile using all three {\sl Chandra} observations in the energy band 0.3 to 2.5\,keV, with point sources excluded. The surface brightness profile from 0.0656\,arcmin (0.319\,kpc) to 1.64\,arcmin (7.97\,kpc) centred on the nucleus is shown in Figure~\ref{fig:3}. Each annulus contains more than 150 counts to ensure a sufficient signal-to-noise ratio. 
We fit the surface brightness profile with a single $\beta$-model:
\begin{eqnarray}
I(r) = I_0[1+(r/r_c)^2]^{(-3\beta+1/2)}.
\end{eqnarray}
The central surface brightness $I_0$, core radius $r_c$ and $\beta$ are the fit parameters. A constant background was also included in the fit. The best-fit parameters of the $\beta$-model are: $r_c$ = $0.92^{+0.15}_{-0.14}$\,kpc and $\beta$ = $0.41^{+0.03}_{-0.02}$.

\begin{figure*}
  \begin{tabular}{c}
    \includegraphics[width=1.9\columnwidth]{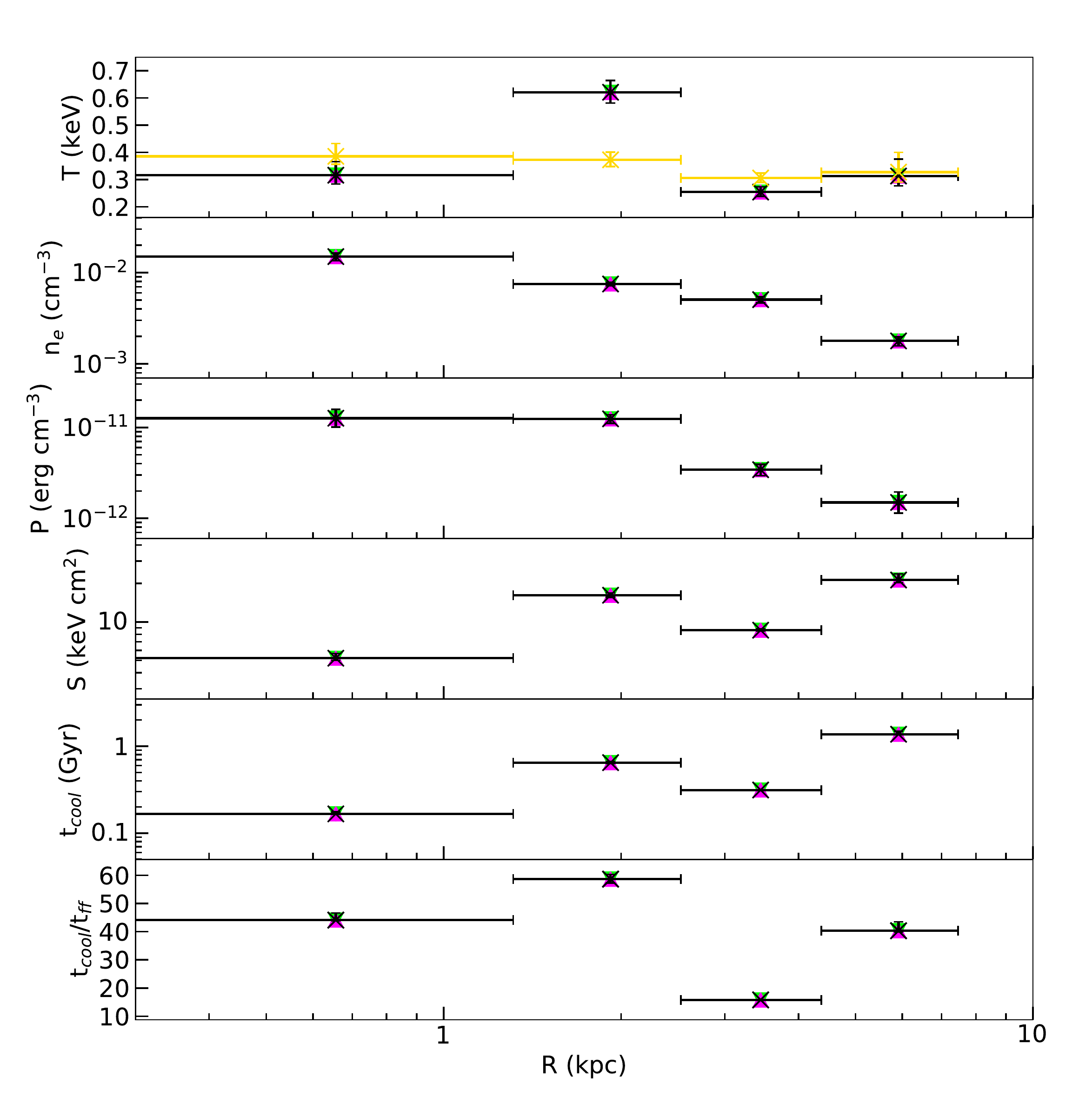} \\
  \end{tabular}
  \caption{Yellow crosses: projected temperature profile. Black crosses: deprojected radial profiles for the hot gas, extracted and fitted in annuli with more than 500 counts. Top to bottom: temperature, electron density, pressure, entropy, cooling time and the ratio of cooling time to free fall time.  Magenta (green) triangles present fits after increasing (decreasing) the normalizations of the LMXBs, ABs and CVs components by 20 per cent.}
  \label{fig:4}
\end{figure*}

Figure~\ref{fig:4} shows the deprojected azimuthally averaged radial profiles for gas temperature, electron density, pressure, entropy, and cooling time. Electron densities are calculated from the normalizations in the {\tt\string apec} model:
\begin{eqnarray}
norm = \frac{10^{-14}}{4\pi[D_A(1+z)]^2}n_en_HV,
\end{eqnarray}
where $V$ is the volume of the concentric spherical shell; $n_e$ and $n_H$ are the electron and proton densities, respectively, for which we assume $n_e/n_H$ = 1.2.  
From the temperature and density profiles, we derive the gas pressure, cooling time and entropy profiles.

The gas pressure $P$ is deduced from:
\begin{eqnarray}
P = 2n_HkT
\end{eqnarray}
The three-dimensional gas entropy is derived from:
\begin{eqnarray}
S = kTn_e^{-2/3}. 
\end{eqnarray}
The cooling time for the gas is calculated as:
\begin{eqnarray}
t_{\rm cool} = \frac{3P}{2n_en_H\Lambda(T,Z)}
\end{eqnarray}
where $\Lambda(T,Z)$ is the cooling function.
 
As shown in Figure~\ref{fig:4}, the temperature of the hot ISM is $\sim$0.3\,keV. There is a temperature jump at $\sim$2\,kpc, indicating possible shock fronts as discussed in \S4.1.
The central density is $\sim$0.014\,cm$^{-3}$. The ISM within $\sim$4.5\,kpc has a cooling timescale less than 1\,Gyr, much shorter than the galaxy age of $\sim$11.7\,Gyr \citep{2010MNRAS.408...97K}. The entropy of the ISM is $\lesssim$20\,keV\,cm$^2$. We classify NGC 4477 as a cool-core galaxy. %


\subsection{Cavities}
\label{sec:3.2}

\begin{figure}
  \begin{tabular}{l}
    \includegraphics[width=0.95\columnwidth]{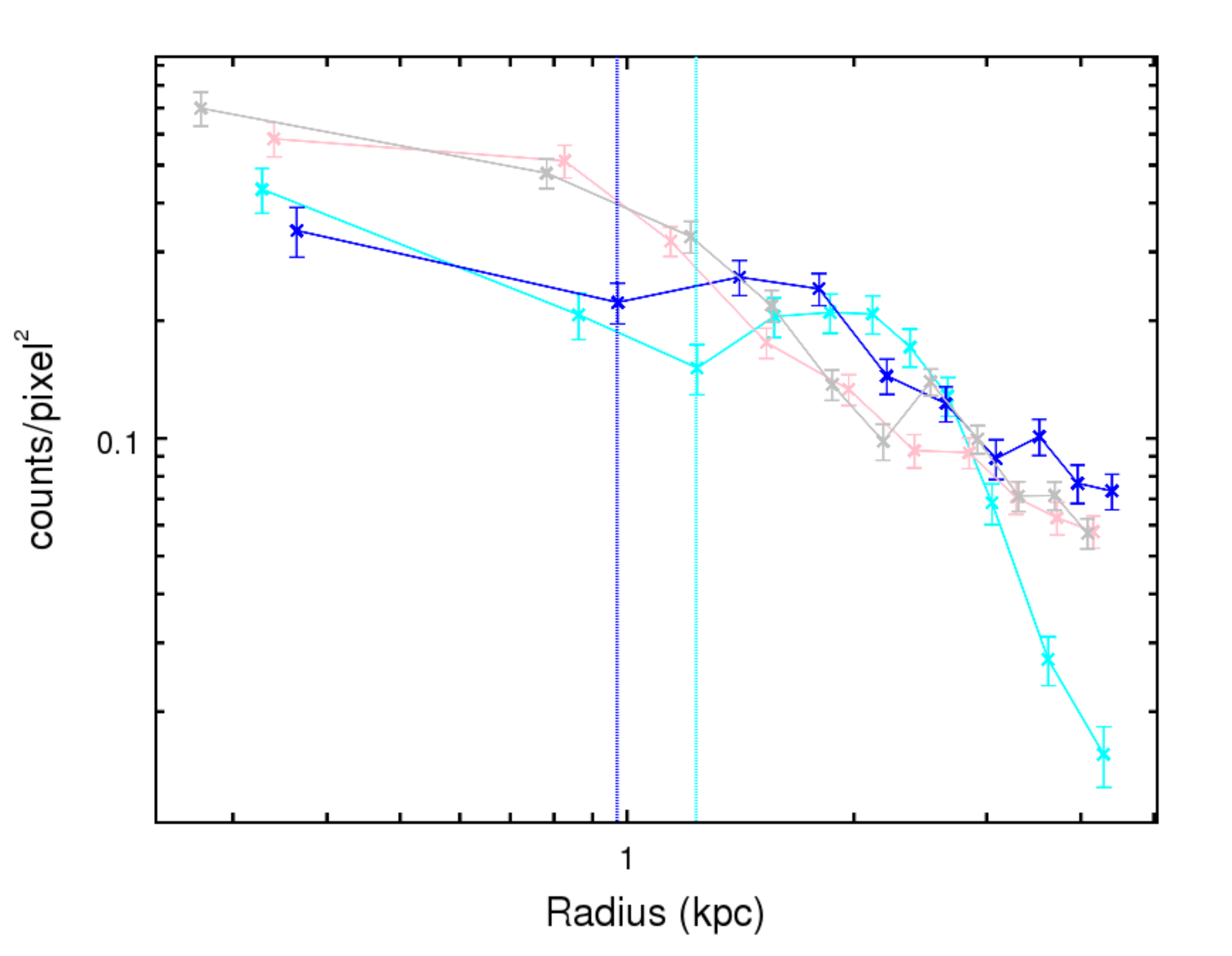}\\
    \includegraphics[width=1\columnwidth]{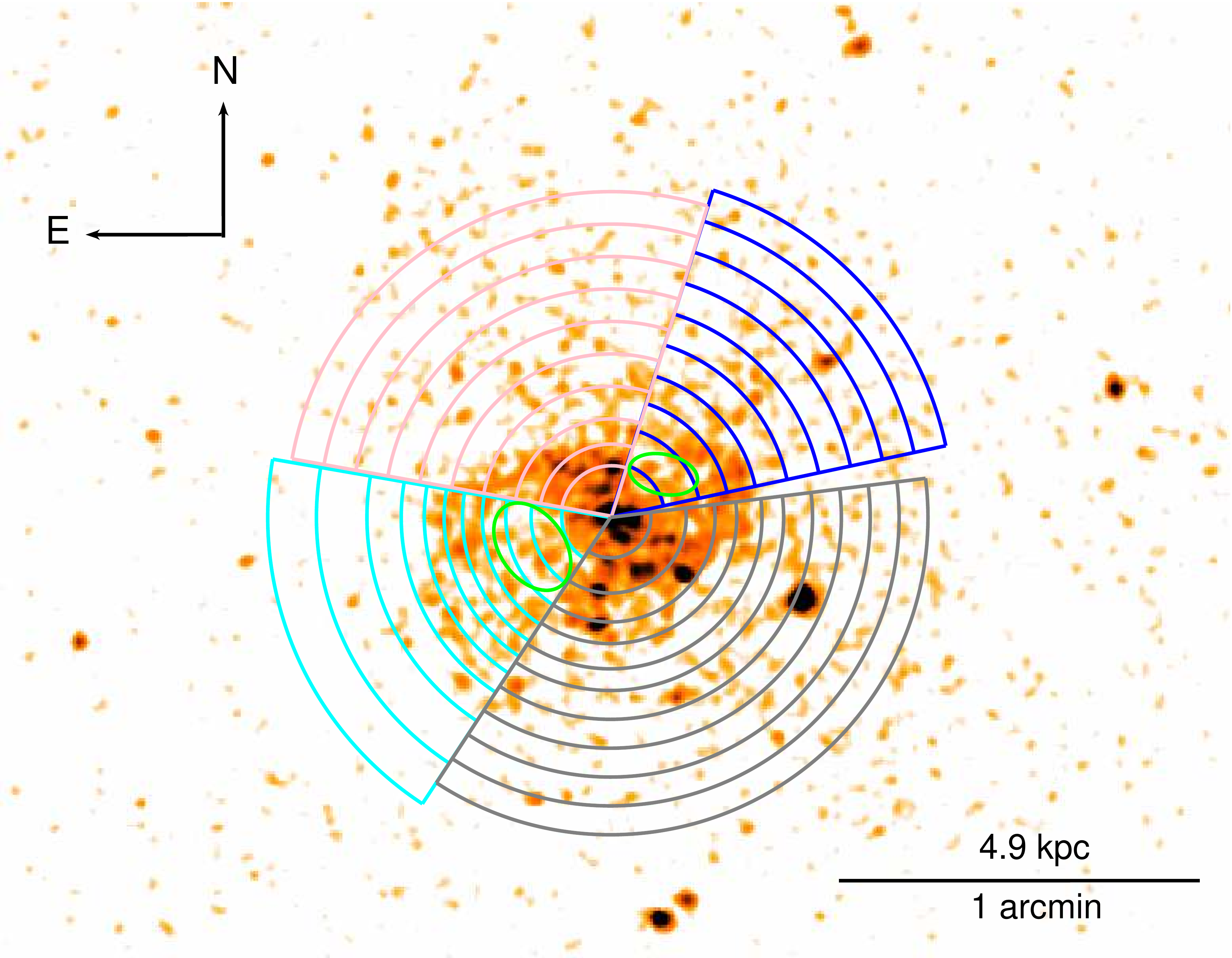}
  \end{tabular}
  \caption{Top: surface brightness profiles of sectors in different directions in the 0.3--2.5\,keV energy band, with 1$\sigma$ errors. Point sources are excluded and blank-sky background has been subtracted.  The cavities coincide with the dips around 1--2\,kpc of the blue and cyan profiles. Dotted lines point out the position of the cavity centres. Bottom: corresponding annuli from which the surface brightness profiles were extracted. Cyan: southeast sector. Blue: northwest sector. Gray: southwest sector. Pink: northeast sector. Cavities are indicated in green.} 
  \label{fig:5} 
\end{figure}
The {\sl Chandra} X-ray image of NGC 4477 reveals a pair of cavities in the hot halo (see Figure~\ref{fig:2}). The detection of the cavities is significant (see \S2.1).
We generate surface brightness profiles of the gas in the 0.3--2.5\,keV energy band along annular sectors in four different directions (Figure~\ref{fig:5}). Two dips can be seen at 0.9-2\,kpc in the surface brightness profiles for the southeast and northwest sectors, consistent with the positions of the cavities on the images. To further highlight the X-ray cavities, we divide the {\sl Chandra} image (Figure~\ref{fig:2}-left) by a double-$\beta$ profile model to obtain a residual image as shown in Figure~\ref{fig:2}-right.

We list the properties of the cavities in Table~\ref{tab:2}. We approximate the cavity shapes by ellipses with a semi-major axis a and a semi-minor axis b. Their volumes are calculated assuming a prolate spheroid. Following \citet{2004ApJ...607..800B}, the upper and lower limits of the volumes are calculated assuming either oblate or prolate symmetry with an a/b ratio of 2.83 (the most eccentric cavity observed in \citep{2004ApJ...607..800B}).
Based on the temperature profile in Figure~\ref{fig:4}, we adopt a temperature $kT$ $\sim 0.3$\,keV for gas around the cavities. The sound speed is calculated as $c_s =  (\gamma kT /m_p\mu)^{1/2}$, 
where $\mu$ = 0.6, $\gamma$ = 5/3, and $m_p$ is the proton mass. In this case $c_s$ is $\sim$282\,km\,s$^{-1}$.
We estimate the velocity of the rising bubbles following \citet{2001ApJ...554..261C}
\begin{eqnarray}
v=\sqrt{g\frac{2V}{SC}}
\end{eqnarray}
where $g$ is the gravitational acceleration, $V$ is the cavity volume, $S$ is the cavity cross section, and $C\approx0.75$ the drag coefficient. We adopt a stellar velocity dispersion of $\sigma=170$\,km s$^{-1}$ for NGC~4477 \citep{2013MNRAS.432.1862C} to calculate $g={2\sigma^2}/r$, where $r$ is the distance from the cavity centre to the nucleus, which is approximately $r\sim1$\,kpc for both cavities. We estimate that the bubbles rise at $v\approx250$\,km s$^{-1}$, roughly 90 per cent of the sound speed. Note that the distances from the cavities to the AGN are only about the diameter of the cavities, indicating that the cavities may have been inflated by a recent AGN outburst. 
We assume the age of the cavities is similar to the rise time $t_{\rm rise}\sim r/v$$\sim$ 4\,Myr. This is much shorter than the cooling time of the central gas of $t_{\rm cool}\lesssim 1$\,Gyr. The enthalpy of the cavities $H$ is the minimum energy required to inflate the cavities, computed by 4$P V$, where $P$ is the pressure at the location of the cavity centre and is obtained from the deprojected radial profiles, and $V$ is the cavity volume. The total enthalpy of the two cavities is $\sim$$10^{54}$\,erg, smaller than the cavities in other systems by at least two orders of magnitude, e.g., 
MS0735.6+7421 \citep{2005Natur.433...45M}, NGC 1399 \citep{2017ApJ...847...94S}, and M84 \citep{2008ApJ...686..911F}.


No radio emission has been detected at the position of the X-ray cavities from previous studies with a 5$\sigma$ flux upper limit of approximately 1.5\,mJy at 1.4\,GHz\footnote{The VLA FIRST survey gives a 5$\sigma$ upper limit of 0.72\,mJy/beam \citep{2017MNRAS.464.1029N}. The diameter of each cavity is approximately twice as large as the beam size, resulting in an upper limit of 1.44\,mJy.}, making them ``ghost bubbles". We assume the spectral index of the radio lobes in NGC 4477 to be $\alpha\sim0.65$, similar to that of M84 \citep{2004ApJ...607..800B} and NGC 4649 \citep{2010MNRAS.404..180D}, which are also individual galaxies; its radio flux is integrated from 10 MHz to 10GHz. We estimate that the cavities have a flux of 80\,$\mu$Jy at 1.4\,GHz, assuming that the total internal pressure equals to the sum of the pressure of the relativistic particles and the magnetic field pressure. However, it was found that the minimum pressure of the radiating particles and the magnetic field generally falls short of the required external pressure (\citealt{2005MNRAS.364.1343D}; \citealt{2008MNRAS.384.1344J}). Therefore, the synchrotron flux of the cavities is likely to be lower than our estimate. It is not surprising that its radio emission cannot be detected with existing observations.

\begin{table}
	\centering
	\caption{Properties of the two X-ray cavities}
	\label{tab:2}
	\begin{tabular}[width=\columnwidth]{cccccc}
		\hline
		ID & {$a^a$} & {$b^b$} & {$r^c$} & {$H^d$} & {$t_{\rm rise}$$^e$} \\
           & (kpc) & (kpc) & (kpc) & ($10^{53}$erg) & (Myr)\\
		\hline
		SE & 0.65 & 0.43 & 1.1 & 7.6$^{+7.0}_{-2.2}$ & 3.8$^{+1.7}_{-0.4}$ \\ 
        NW & 0.47 & 0.27 & 0.9 & 2.1$^{+1.4}_{-1.4}$ & 3.6$^{+1.1}_{-1.1}$ \\ 
		\hline
     \multicolumn{6}{p{2.75in}}{$^a$ Semi-major axis.}\\
     \multicolumn{6}{p{2.75in}}{$^b$ Semi-minor axis.}\\
     \multicolumn{6}{p{2.75in}}{$^c$ Distance from the nucleus.}\\
     \multicolumn{6}{p{2.75in}}{$^d$ Energy required to produce the cavity.}\\
     \multicolumn{6}{p{2.75in}}{$^e$ Rise time of the cavity from the central nucleus.}
     \end{tabular}
\end{table}

\section{Discussion}
\label{sec:4}
The ISM of NGC 4477 has a low central temperature $\sim$0.3 keV, a short central cooling time $\lesssim$1\,Gyr and a central entropy of $\lesssim$20\,keV\,cm$^2$. 
In this paper we identify a pair of X-ray cavities with 1.3\,kpc and 0.9\,kpc diameters residing along the east-west axis.
Below we examine the ISM substructures and discuss the AGN feedback mechanism in this small lenticular galaxy.

\subsection{Edges}
\label{sec:4.1}
\begin{figure}
    \includegraphics[width=\columnwidth]{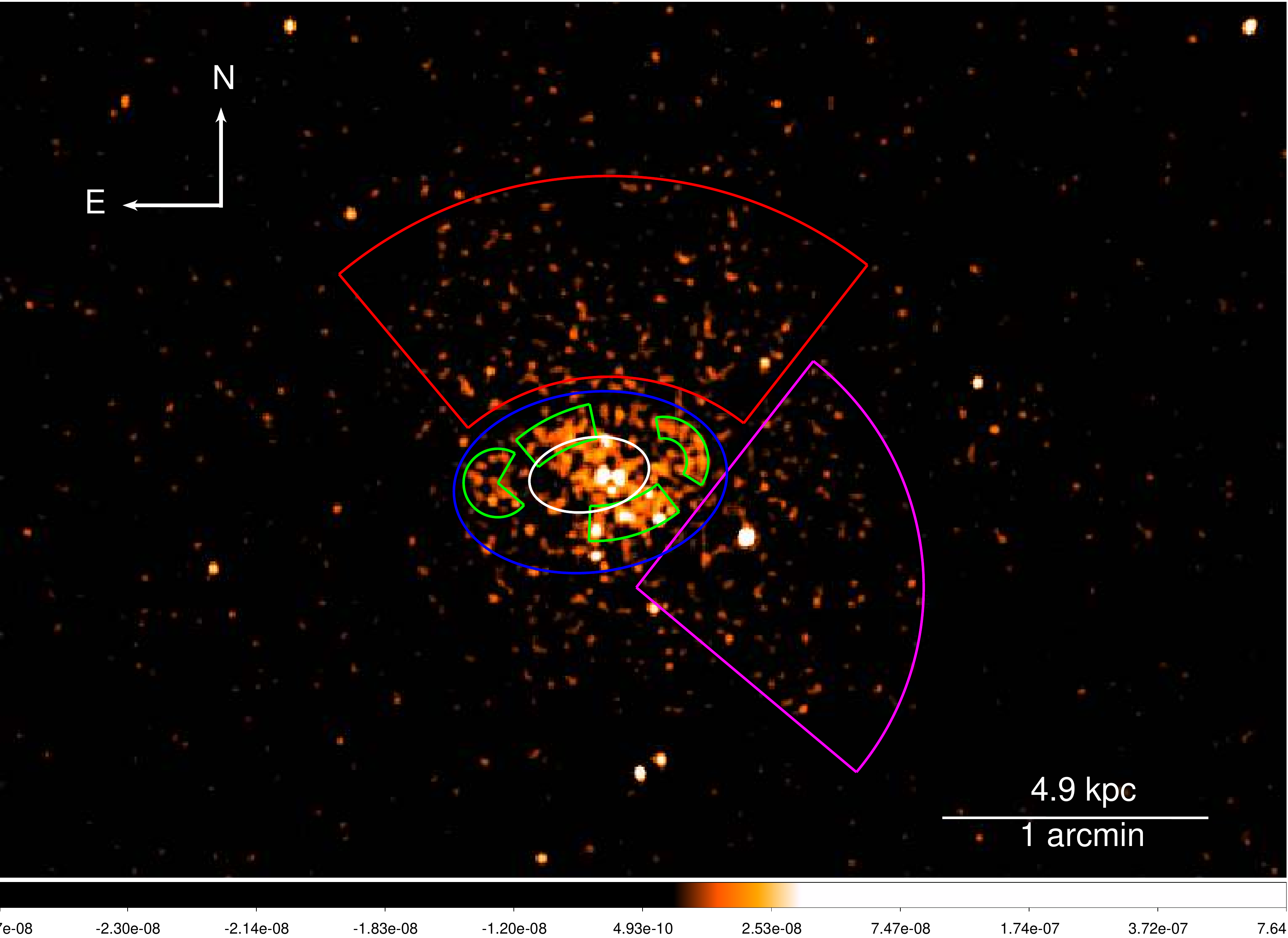} \\
    \caption{ White, green, red, and magenta contours denote regions for which we measured gas temperature. White: centre. Green: edges. Red: northern gas. Magenta: southwestern gas. We also fit the spectra in a region enclosed by the blue contour, with the ``centre" and ``edges" regions excluded. This region is labeled as ``edges' surroundings" in Table~\ref{tab:3}.}
    \label{fig:6}
\end{figure}

Thw Chandra image reveals sharp surface brightness edges outside the cavities, indicating possible shock fronts as shown in Figure~\ref{fig:2}. The gas temperature profile has a jump at $r\sim 2$\,kpc (see the top panel in Figure~\ref{fig:4}), coincident with the edge positions, providing additional support for shock fronts. 
To better characterize the possible shock features, we perform spectrum analyses in the 0.5--7.0\,keV energy band for the central region, edges, the ambient region and two outer more diffuse regions. These regions are indicated in Figure~\ref{fig:6} and the best-fit temperatures are listed in Table~\ref{tab:3}.
The temperature of the edges is at least 60 per cent (0.2 keV) higher than other regions, suggesting shock heated gas. 
The presence of shocks would suggest a relatively recent AGN outburst in NGC 4477, consistent with the short rising time for the bubbles (\S3.2). The AGN outbursts may deposit energy into the ambient gas through both cavities and shocks. In this scenario, the internal pressure of the cavities would exceed the external pressure, invalidating the assumption we made in deriving its radio flux (\S3.2). Then again, we may have overestimated its emission by neglecting non-radiating particles.
Deeper observations would allow us to characterize the gas density jumps across the shock fronts and the energy in shocks.

\begin{table}
	\centering
	\caption{Temperatures of regions around the edges as marked in Figure~\ref{fig:6}}
	\label{tab:3}
	\begin{tabular}[width={3in}]{cc}
		\hline
		Region & Temperature \\
           &  (keV) \\
        \hline
        Centre & $0.37^{+0.12}_{-0.05}$\\
        Edges & $0.61^{+0.07}_{-0.06}$ \\
        Edges' Surroundings & $0.35^{+0.05}_{-0.04}$ \\
        Northern Gas & $0.33^{+0.05}_{-0.03}$ \\
        Southwestern Gas & $0.30^{+0.03}_{-0.03}$ \\
       \hline
     \end{tabular}
\end{table}

\subsection{The Balance Between Cavity Heating and Cooling}
\label{sec:4.2}
We measure a bolometric X-ray luminosity of $L_{\rm bol}\sim 5.3$$\times 10^{39}$\,erg\,s$^{-1}$ for the hot gas within a radius of $\sim$4.5\,kpc ($t_{\rm cool}$ $\lesssim1$\,Gyr). 
We assume 100 per cent of the cavities' energy goes into heating the gas. 
The total enthalpy of the two cavities is $\sim$$10^{54}$\,ergs with an age of $\sim$4\,Myr, 
corresponding to a cavity power of $P_{\rm cav}$ = $8.3^{+6.6}_{-2.3} \times 10^{39}$\,erg\,s$^{-1}$, comparable to the level of cooling. Furthermore, as we noted in \S4.1, AGN outbursts could also heat the ISM through shocks. Therefore the outburst energy is sufficient to reheat cooling gas in NGC~4477.

\begin{figure}
  \begin{tabular}{l}
    \includegraphics[width=\columnwidth]{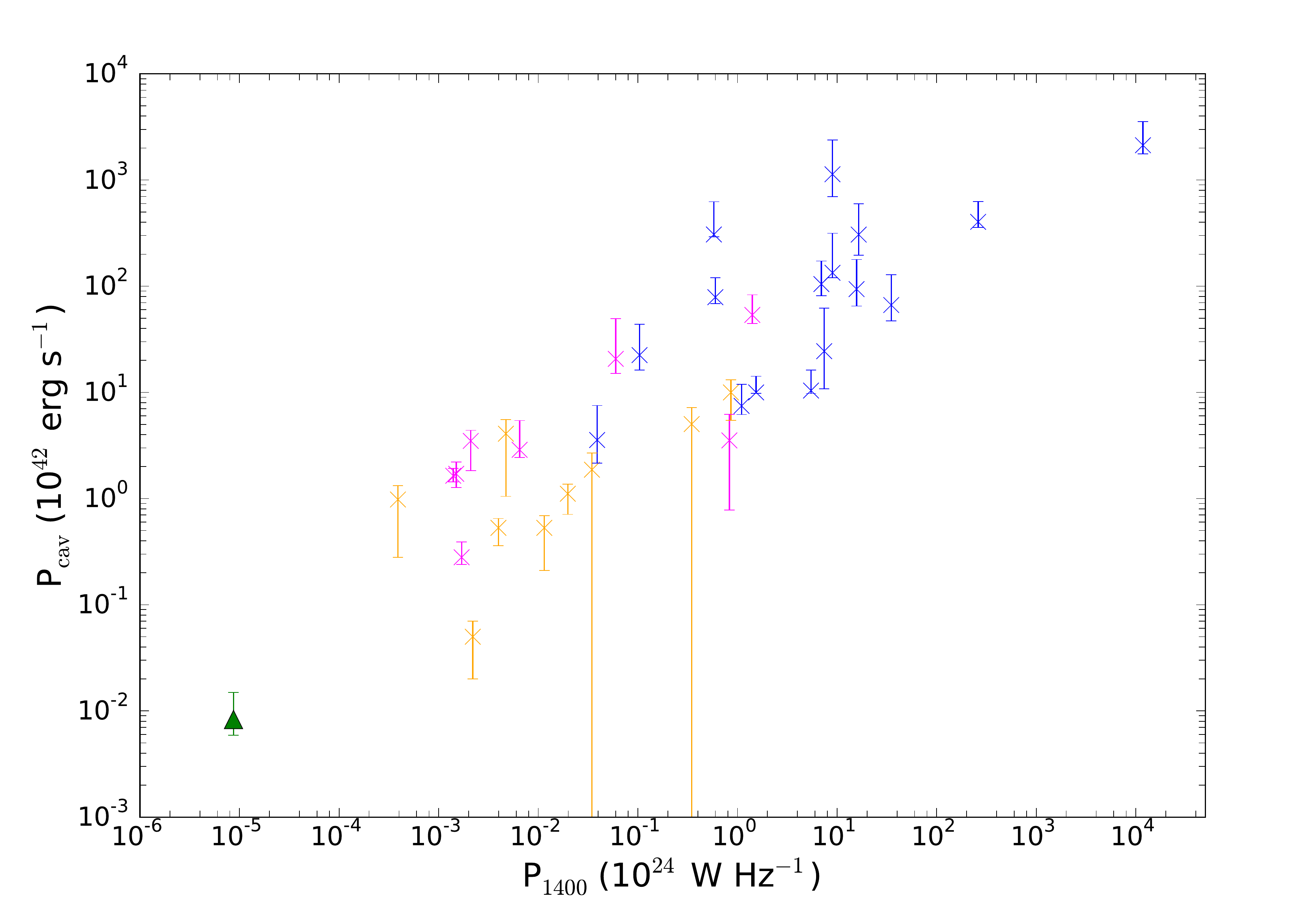} \\
  \end{tabular}
  \caption{Scaling relations between cavity power and the total radio power of the system at 1.4\,GHz. Magenta: \citet{2011ApJ...735...11O}; Blue: \citet{2004ApJ...607..800B}; Orange: \citet{2010ApJ...720.1066C}. The green triangle represents NGC 4477.}
  \label{fig:7}
\end{figure}

An approximate linear relation between the logarithm of cavity power and total radio power from the nucleus and the cavity-filling plasma in clusters, groups and individual elliptical galaxies has been well established (\citealt{2004ApJ...607..800B,2010ApJ...720.1066C,2011ApJ...735...11O}).
NGC~4477 is a lenticular galaxy with cavity power smaller than any other cool core system. Its cavity power $P_{\rm cav}$ = $8.3^{+6.6}_{-2.3} \times 10^{39}$\,erg\,s$^{-1}$ and 1.4\,GHz radio power $P_{1400}$ = 8.7$\times 10^{18}$\,W\,Hz$^{-1}$ \citep{2001ApJS..133...77H} are in agreement with this relation as shown in Figure~\ref{fig:7}.
Cavity heating may be a common heating mechanism in the hot gas corona of different systems spanning a wide range of masses, sizes, and environments. It is desirable to investigate more small systems like NGC~4477.

\subsection{The Central Sources}
\label{sec:4.3}
The {\sl Chandra} X-ray images (Figures~\ref{fig:2}, \ref{fig:5} and \ref{fig:6}) show two bright point sources (S1 and S2) near the galaxy centre. We wish to determine their nature and identify which one is the nucleus. NGC 4477 is a Seyfert 2 galaxy and the nucleus may be heavily absorbed. We modeled the X-ray emission from the two point sources with a  {\tt\string wabs $\times$ zpowerlw $\times$ zwabs} model, using circular regions with a radius $\sim$1.8\,arcsec and a background obtained from a local annular region. The {\tt\string wabs} model accounts for the galactic absorption and the {\tt\string zwabs} accounts for the intrinsic absorption. The photon indices were fixed at 1.8 (typical for intrinsic AGN spectra, see \citealt{2010ApJ...720..368X}). The best-fit local absorption column densities are 5.4$\times 10^{20}$\,cm$^{-2}$ (S1) and 1.5$\times 10^{21}$\,cm$^{-2}$ (S2). The extrapolated X-ray luminosities of the two nuclei in the 2.0--10.0\,keV energy band are 4.6$^{+0.6}_{-0.5} \times 10^{38}$\,erg\,s$^{-1}$ (S1) and 3.6$^{+0.5}_{-0.5} \times 10^{38}$\,erg\,s$^{-1}$ (S2). Our results are approximately consistent with the measurements of \citet{2006A&A...446..459C} using {\sl XMM-Newton} observation. They measured the nucleus luminosity in a larger region with a radius $\sim$15\,arcsec, which contains both point sources, and gave an upper limit for absorption column density of 2$\times 10^{22}$\,cm$^{-2}$, the hard X-ray photon index of 1.9 $\pm$ 0.3, and the X-ray luminosity in the 2.0--10.0\,keV energy band of 4.0$\times 10^{39}$\,erg\,s$^{-1}$ (converted for the distance used in this paper). 
Based on their X-ray luminosities, these two point sources can be either AGNs or LMXBs.

\begin{figure}
  \begin{tabular}{l}
    \includegraphics[width=\columnwidth]{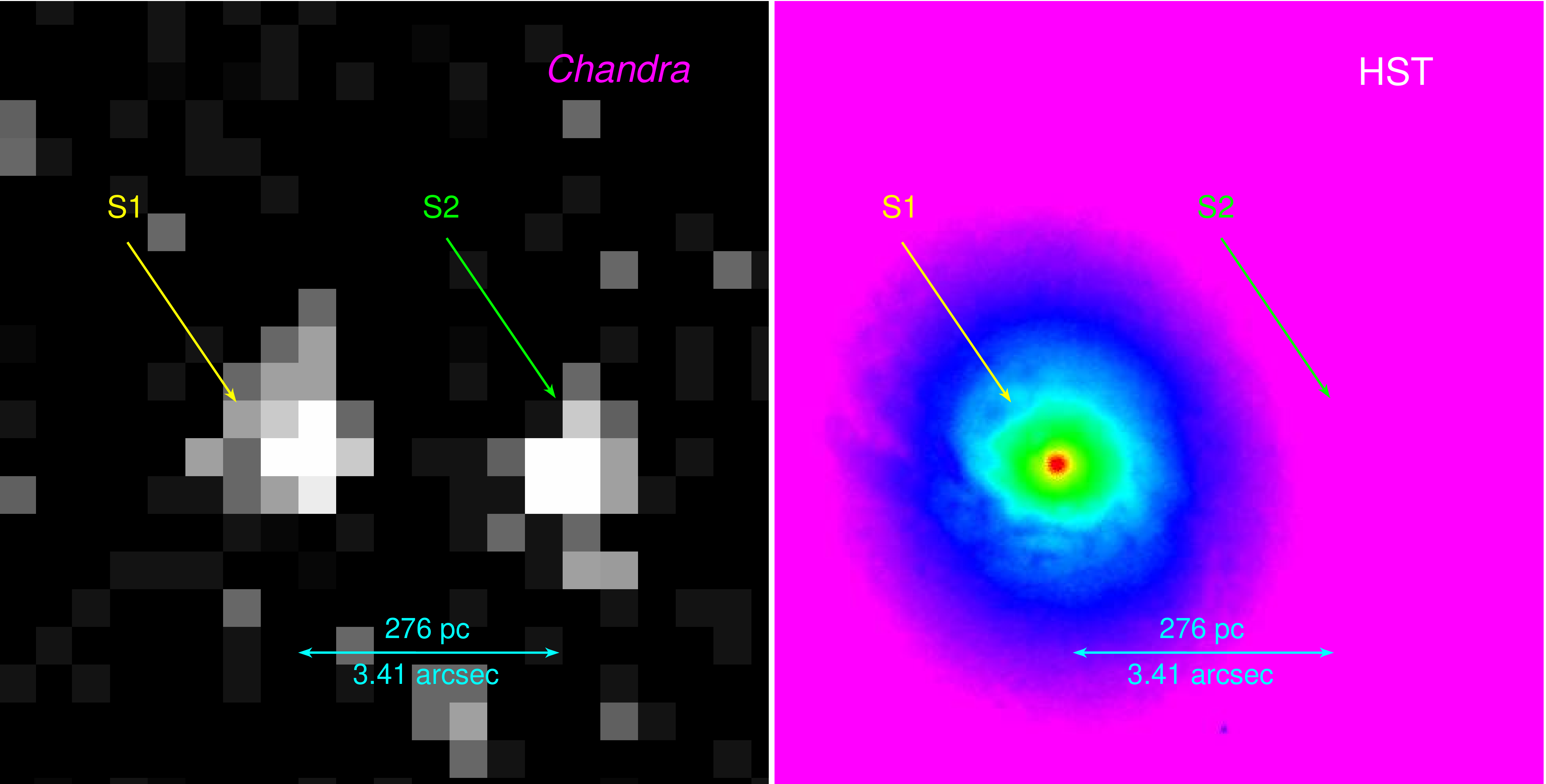} \\
  \end{tabular}
  \caption{{\sl Chandra} (left) and {\sl HST} WFC3 (right) images of the two X-ray point sources S1 and S2. S1 has an optical counterpart while S2 does not.} 
  \label{fig:8}
\end{figure}

Radio emission from S1 was detected at 6cm by \citet{2001ApJS..133...77H}, while no radio emission from S2 was detected. S1 could still be discerned from the {\sl Chandra} X-ray image above 6 keV while S2 could not. 
As shown in the {\sl HST} image (Figure~\ref{fig:8}-right), S1 has an optical counterpart while S2 does not. S1 is also closer to the midpoint between the locations of the two cavities. 
In addition, the relation between the [OIII] and H$_\alpha$ luminosities of S1 taken from \citet{1997ApJS..112..315H} agrees with that for AGNs found by \citet{2006A&A...455..173P}. For the reasons above, we identify S1 as the galaxy nucleus which has inflated the cavities. 
Based on the number counts--flux relation from {\sl Chandra} Deep Field-South \citep{2012ApJ...752...46L}, the binomial probability of finding two such low luminosity AGNs at a distance of 0.001 deg, which is the approximate distance between the two point sources, is about 0.06 per cent. 
Therefore, S2 could be a background AGN but is more likely an LMXB.

\subsection{Cold Gas Contents}
\label{sec:4.4}

\begin{figure}
  \begin{tabular}{l}
    \includegraphics[width=0.5\columnwidth]{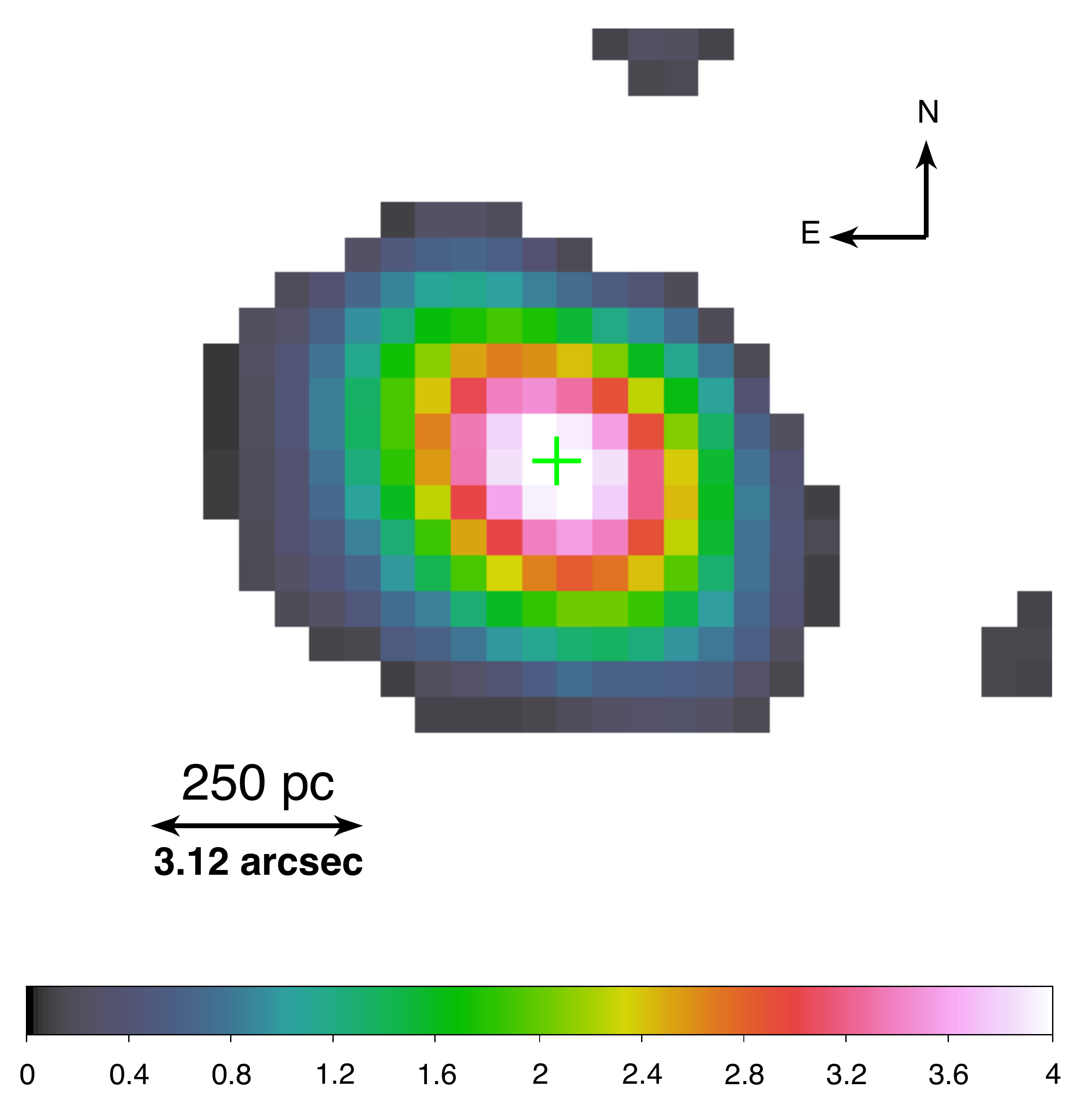}
    \includegraphics[width=0.5\columnwidth]{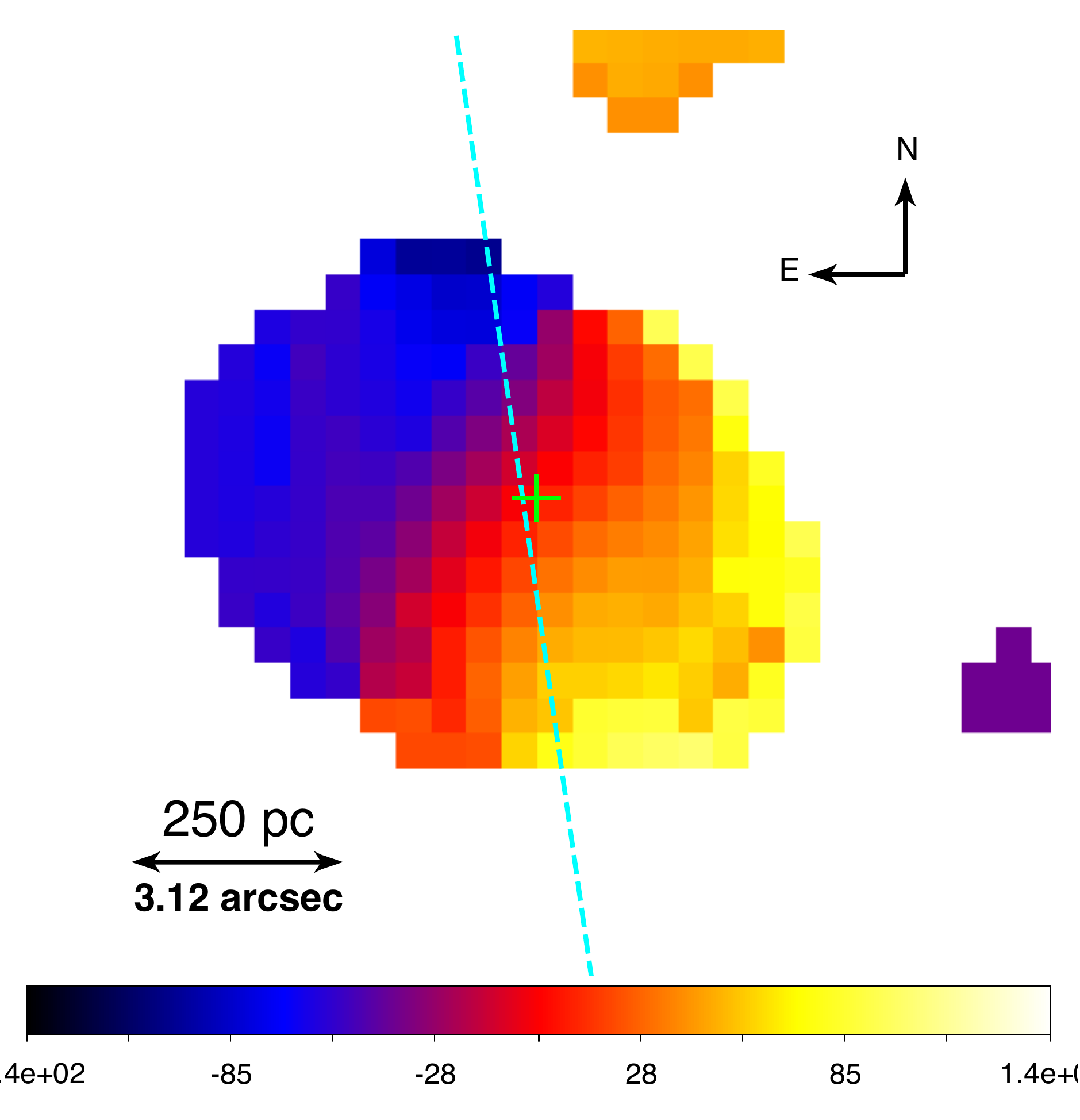}
   \\
  \end{tabular}
  \caption{IRAM CO(1-0) properties of NGC~4477 taken from \citet{2011MNRAS.410.1197C}. {\it left}: integrated intensity maps in units of Jy\,beam$^{-1}$\,km\,s$^{-1}$. The beam size is 3.3 $\times$ 2.6 arcsec$^2$. {\it right}: mean velocity maps in units of km\,s$^{-1}$, relative to the radial velocity of NGC~4477 taken from NED. The rotation axis of cold gas is misaligned with that of the stellar disk, where the cyan dashed line indicates the alignment of the stellar disk as shown in Figure~\ref{fig:1}. The blue cross indicates the position of the nucleus.} 
  \label{fig:9}
\end{figure}

Feedback from the central supermassive black hole is by far the most viable solution to the cooling problem. Hot plasma in the centres of cool core systems radiate strongly in X-rays. Plasma with a sufficiently small cooling time to free fall time ratio, $t_{\rm cool}/t_{\rm ff}$, is thermally unstable and may condense into cold gas. Such cold gas can accrete onto the black hole and ignite AGN outbursts \citep[e.g.,][]{2012ApJ...746...94G,2012MNRAS.419.3319M}. This paradigm is supported by the detection of cold gas in a growing number of cool core systems \citep[e.g.,][]{2011scgg.conf...11E,2017ApJ...842...84D}. These systems usually have a minimum $t_{\rm cool}/t_{\rm ff}$ ratio smaller than 20 \citep{2017arXiv170400011H}. 
A significant amount of cold molecular gas has been detected in NGC~4477 with $M_{H_2} = 0.394 \times 10^8 M_{\sun}$ and $M_{H_I} < 0.089 \times 10^8 M_{\sun}$ \citep{{2011MNRAS.414..940Y}}. 
We calculate $t_{\rm ff}=\sqrt{2r/g}=r/\sigma$ to obtain $t_{\rm cool}/t_{\rm ff}$ as a function of radius as shown in Figure~\ref{fig:4}. The $t_{\rm cool}/t_{\rm ff}$ ratio is not small enough to set off condensation. 
As shown in the Figure~9 taken from \citet{2011MNRAS.410.1197C}, cold gas in NGC~4477 is in a shape of a regular disk with ordered rotation within a radius of $r<0.4$\,kpc. In contrast, thermally unstable plasma tends to be filamentary and is often spatially associated with AGN bubbles (e.g., \citealt{2016MNRAS.458.3134R}; \citealt{2017arXiv171004664P}). The mass of the cold gas in NGC~4477 ($3.9 \times 10^7 M_{\sun}$, mainly in $M_{H_2}$) is approximately 70 per cent that of the hot gas within the cool core ($5.8 \times 10^7 M_{\sun}$). NGC~4477 has a cooling rate of $\dot{M}_{\rm cool}=0.03\,M_{\sun}/{\rm yr}$ at the galaxy centre, where cold gas is found and $t_{\rm cool}<0.2$\,Gyr. It would take 1.6\,Gyr to form the observed amount of cold gas via radiative cooling. Cold gas at the centre of NGC~4477 is unlikely to be part of the feedback loop. 

Finally we note that the rotating disk of cold gas is misaligned with the stellar disk of the galaxy (see Figures~\ref{fig:1} and \ref{fig:9}), suggesting that the cold gas in NGC~4477 is accreted externally rather than originating from the internal stellar mass loss. The most plausible scenario is that NGC~4477 has experienced the merger of a gas-rich dwarf galaxy that has dumped its cold gas at the centre of NGC~4477. 

\section{Conclusion}
\label{sec:5}
In this paper we present a pioneering investigation of X-ray cavities in a lenticular galaxy NGC 4477. We study the gas distribution and the interaction between the AGN and the ISM. We find that:

1. The central gas temperature of NGC~4477 is $\sim$0.3\,keV. The central cooling time and entropy are $\lesssim$1\,Gyr and $\lesssim$20\,keV\,cm$^2$, respectively, suggesting a possible cool core.

2. A pair of cavities lie roughly symmetric and collinear at $\sim$1.1\,kpc and 0.9\,kpc from the central AGN. Their ages are $\sim$4\,Myr and the energy required to inflate the cavities is $\sim$$10^{54}$\,ergs. The cavity power in NGC 4477 is comparable to the radiative cooling of the hot ISM. 

3. We detect possible shock fronts just outside the cavities. The temperature distribution indicates an increase at the position of the bright edges relative to the ambient gas. But deeper observations are required to confirm the possible shock fronts. 

4. NGC 4477 is a lenticular galaxy; its size and mass are small when compared to other cool core systems. It follows the scaling relation between the cavity power and the radio power calibrated for galaxy clusters, groups and individual elliptical galaxies. 

5. Two X-ray bright point sources are visible at the galaxy centre. We determine that the eastern source, S1, is the nucleus of NGC 4477 while the western source, S2, is more likely to be a LMXB. 

6. NGC~4477 contains a significant amount ($4\times10^7$\,M$_{\sun}$) of cold gas, forming a regular disk with ordered motion. Such cold gas may be accreted externally and not related to the AGN feedback. 

\section*{Acknowledgements}

The scientific results are based on observations made by the {\sl Chandra} X-ray Observatory. This work is supported by the National Science Foundation of China under grant J1210039.
We thank Paul Nulsen, Ewan O'Sullivan, and Larry David for their inspiring discussions.  
We thank Alison Crocker for providing us the cold gas image of NGC~4477 in Figure~\ref{fig:9}. We also thank the referee for their insightful comments. YL thanks Felipe Andrade-Santos and Scott Randall for helpful advice on the spectral analysis, and thanks Zhiyuan Li for his comments on the paper. YL appreciates the warm hospitality of Matthew Ashby and Jonathan McDowell during her visit to the Harvard-Smithsonian Center for Astrophysics, and acknowledges support from Top-notch Academic Programs Project of Jiangsu Higher Education Institutions.

\bsp
\label{lastpage}

\begin{thebibliography}{99}
\bibitem[Asplund et al.(2006)]{2006NuPhA.777....1A} Asplund, M., Grevesse, N., \& Jacques Sauval, A.\ 2006, Nuclear Physics A, 777, 1
\bibitem[Beifiori et al.(2012)]{2012MNRAS.419.2497B} Beifiori, A., Courteau, S., Corsini, E.~M., \& Zhu, Y.\ 2012, \mnras, 419, 2497
\bibitem[B{\^i}rzan et al.(2004)]{2004ApJ...607..800B} B{\^i}rzan, L., Rafferty, D.~A., McNamara, B.~R., Wise, M.~W., \& Nulsen, P.~E.~J.\ 2004, \apj, 607, 800
\bibitem[Boroson et al.(2011)]{2011ApJ...729...12B} Boroson, B., Kim, D.-W., \& Fabbiano, G.\ 2011, \apj, 729, 12
\bibitem[Cappi et al.(2006)]{2006A&A...446..459C} Cappi, M., Panessa, F., Bassani, L., et al.\ 2006, \aap, 446, 459
\bibitem[Cappellari et al.(2013)]{2013MNRAS.432.1862C} Cappellari, M., McDermid, R.~M., Alatalo, K., et al.\ 2013, \mnras, 432, 1862
\bibitem[Cavagnolo et al.(2010)]{2010ApJ...720.1066C} Cavagnolo, K.~W., McNamara, B.~R., Nulsen, P.~E.~J., et al.\ 2010, \apj, 720, 1066
\bibitem[Churazov et al.(2001)]{2001ApJ...554..261C} Churazov, E., Br{\"u}ggen, M., Kaiser, C.~R., B{\"o}hringer, H., \& Forman, W.\ 2001, \apj, 554, 261
\bibitem[Crocker et al.(2011)]{2011MNRAS.410.1197C} Crocker, A.~F., Bureau, M., Young, L.~M., \& Combes, F.\ 2011, \mnras, 410, 1197
\bibitem[David et al.(2006)]{2006ApJ...653..207D} David, L.~P., Jones, C., Forman, W., Vargas, I.~M., \& Nulsen, P.\ 2006, \apj, 653, 207
\bibitem[David et al.(2017)]{2017ApJ...842...84D} David, L.~P., Vrtilek, J., O'Sullivan, E., et al.\ 2017, \apj, 842, 84
\bibitem[Davis et al.(2014)]{2014MNRAS.444.3427D} Davis, T.~A., Young, L.~M., Crocker, A.~F., et al.\ 2014, \mnras, 444, 3427
\bibitem[Dong et al.(2010)]{2010ApJ...712..883D} Dong, R., Rasmussen, J., \& Mulchaey, J.~S.\ 2010, \apj, 712, 883
\bibitem[Dunn, et al.(2005)]{2005MNRAS.364.1343D} Dunn, R.~J.~H., Fabian, A.~C. \& Taylor, G.~B.\ 2005, \mnras, 364, 1343.
\bibitem[Dunn \& Fabian(2006)]{2006MNRAS.373..959D} Dunn, R.~J.~H., \& Fabian, A.~C.\ 2006, \mnras, 373, 959
\bibitem[Dunn et al.(2010)]{2010MNRAS.404..180D} Dunn, R.~J.~H., Allen, S.~W., Taylor, G.~B., et al.\ 2010, \mnras, 404, 180 
\bibitem[Edge(2011)]{2011scgg.conf...11E} Edge, A.\ 2011, Structure in Clusters and Groups of Galaxies in the Chandra Era, 11
\bibitem[Fabian(1994)]{1994ARA&A..32..277F} Fabian, A.~C.\ 1994, \araa, 32, 277
\bibitem[Fabian(2012)]{2012ARA&A..50..455F} Fabian, A.~C.\ 2012, \araa, 50, 455
\bibitem[Finoguenov et al.(2008)]{2008ApJ...686..911F} Finoguenov, A., Ruszkowski, M., Jones, C., et al.\ 2008, \apj, 686, 911-917
\bibitem[Gaspari et al.(2012)]{2012ApJ...746...94G} Gaspari, M., Ruszkowski, M., \& Sharma, P.\ 2012, \apj, 746, 94
\bibitem[Heckman \& Best(2014)]{2014ARA&A..52..589H} Heckman, T.~M., \& Best, P.~N.\ 2014, \araa, 52, 589
\bibitem[Ho et al.(1997)]{1997ApJS..112..315H} Ho, L.~C., Filippenko, A.~V., \& Sargent, W.~L.~W.\ 1997, \apjs, 112, 315
\bibitem[Ho \& Ulvestad(2001)]{2001ApJS..133...77H} Ho, L.~C., \& Ulvestad, J.~S.\ 2001, \apjs, 133, 77
\bibitem[Hogan et al.(2017)]{2017arXiv170400011H} Hogan, M.~T., McNamara, B.~R., Pulido, F., et al.\ 2017, arXiv:1704.00011 
\bibitem[Irwin et al.(2003)]{2003ApJ...587..356I} Irwin, J.~A., Athey, A.~E., \& Bregman, J.~N.\ 2003, \apj, 587, 356 
\bibitem[Jetha, et al.(2008)]{2008MNRAS.384.1344J} Jetha, N.~N., Hardcastle, M.~J., Babul, A., et al.\ 2008, \mnras, 384, 1344.
\bibitem[Kalberla et al.(2005)]{2005A&A...440..775K} Kalberla, P.~M.~W., Burton, W.~B., Hartmann, D., et al.\ 2005, \aap, 440, 775 
\bibitem[Kormendy \& Ho(2013)]{2013ARA&A..51..511K} Kormendy, J., \& Ho, L.~C.\ 2013, \araa, 51, 511
\bibitem[Kuntschner et al.(2010)]{2010MNRAS.408...97K} Kuntschner, H., Emsellem, E., Bacon, R., et al.\ 2010, \mnras, 408, 97
\bibitem[Lehmer et al.(2012)]{2012ApJ...752...46L} Lehmer, B.~D., Xue, Y.~Q., Brandt, W.~N., et al.\ 2012, \apj, 752, 46
\bibitem[McCourt et al.(2012)]{2012MNRAS.419.3319M} McCourt, M., Sharma, P., Quataert, E., \& Parrish, I.~J.\ 2012, \mnras, 419, 3319
\bibitem[McNamara et al.(2005)]{2005Natur.433...45M} McNamara, B.~R., Nulsen, P.~E.~J., Wise, M.~W., et al.\ 2005, \nat, 433, 45
\bibitem[McNamara \& Nulsen(2007)]{2007ARA&A..45..117M} McNamara, B.~R., \& Nulsen, P.~E.~J.\ 2007, \araa, 45, 117
\bibitem[Mei et al.(2007)]{2007ApJ...655..144M} Mei, S., Blakeslee, J.~P., C{\^o}t{\'e}, P., et al.\ 2007, \apj, 655, 144
\bibitem[Nulsen et al.(2009)]{2009AIPC.1201..198N} Nulsen, P., Jones, C., Forman, W., et al.\ 2009, American Institute of Physics Conference Series, 1201, 198
\bibitem[Nyland et al.(2017)]{2017MNRAS.464.1029N} Nyland, K., Young, L.~M., Wrobel, J.~M., et al.\ 2017, \mnras, 464, 1029
\bibitem[O'Sullivan et al.(2011)]{2011ApJ...735...11O} O'Sullivan, E., Giacintucci, S., David, L.~P., et al.\ 2011, \apj, 735, 11
\bibitem[Panessa et al.(2006)]{2006A&A...455..173P} Panessa, F., Bassani, L., Cappi, M., et al.\ 2006, \aap, 455, 173
\bibitem[Pulido et al.(2017)]{2017arXiv171004664P} Pulido, F.~A., McNamara, B.~R., Edge, A.~C., et al.\ 2017, arXiv:1710.04664
\bibitem[Revnivtsev et al.(2008)]{2008A&A...490...37R} Revnivtsev, M., Churazov, E., Sazonov, S., Forman, W., \& Jones, C.\ 2008, \aap, 490, 37
\bibitem[Russell et al.(2016)]{2016MNRAS.458.3134R} Russell, H.~R., McNamara, B.~R., Fabian, A.~C., et al.\ 2016, \mnras, 458, 3134
\bibitem[Shin et al.(2016)]{2016ApJS..227...31S} Shin, J., Woo, J.-H., \& Mulchaey, J.~S.\ 2016, \apjs, 227, 31
\bibitem[Su \& Irwin(2013)]{2013ApJ...766...61S} Su, Y., \& Irwin, J.~A.\ 2013, \apj, 766, 61
\bibitem[Su et al.(2015)]{2015ApJ...806..156S} Su, Y., Irwin, J.~A., White, R.~E., III, \& Cooper, M.~C.\ 2015, \apj, 806, 156
\bibitem[Su et al.(2017)]{2017ApJ...847...94S} Su, Y., Nulsen, P.~E.~J., Kraft, R.~P., et al.\ 2017, \apj, 847, 94
\bibitem[Xue et al.(2010)]{2010ApJ...720..368X} Xue, Y.~Q., Brandt, W.~N., Luo, B., et al.\ 2010, \apj, 720, 368
\bibitem[Young et al.(2011)]{2011MNRAS.414..940Y} Young, L.~M., Bureau, M., Davis, T.~A., et al.\ 2011, \mnras, 414, 940

\end{thebibliography}
\end{document}